\documentclass[aps,preprint,pre,showpacs,eqsecnum]{revtex4-1}
\usepackage{graphicx} 
\usepackage{color}
\definecolor{Green}{rgb}{0,0.5,0}

\usepackage[english]{babel}
\usepackage[utf8]{inputenc}

\usepackage{amsmath,amssymb}

\usepackage{textcomp}

\begin{document}
%%%%%%%%%%%%%%%%%%%%%%%%%%%%%%%%%%%%%%%%%%%%%%%
%%% title %%%%%%%%%%%%%%%%%%%%%%%%%%%%%%%%%%%%%
%%%%%%%%%%%%%%%%%%%%%%%%%%%%%%%%%%%%%%%%%%%%%%%

\title{Reactive-Ion-Etched Graphene Nanoribbons on a Hexagonal Boron Nitride Substrate}
%\author{D. Bischoff$^{1,*}$, T. Krähenmann$^{1}$, S. Dröscher$^{1}$, M. A. Gruner$^{1}$, C. Barraud$^{1}$, T. Ihn$^{1}$, K. Ensslin$^{1}$}
%\affiliation{$^1$Solid State Physics Laboratory, ETH Zurich, 8093 Zurich, Switzerland, $^*$dominikb@phys.ethz.ch}
\author{D. Bischoff}
\email[corresponding author: ]{dominikb@phys.ethz.ch}
%\noaffiliation
\author{T. Krähenmann}
%\noaffiliation
\author{S. Dröscher}
%\noaffiliation
\author{M. A. Gruner}
%\noaffiliation
\author{C. Barraud}
%\noaffiliation
\author{T. Ihn}
%\noaffiliation
\author{K. Ensslin}
%\noaffiliation
\affiliation{Solid State Physics Laboratory, ETH Zurich, 8093 Zurich, Switzerland}
\date{\today}

%%%%%%%%%%%%%%%%%%%%%%%%%%%%%%%%%%%%%%%%%%%%%%%
%%% abstract%%%%%%%%%%%%%%%%%%%%%%%%%%%%%%%%%%%
%%%%%%%%%%%%%%%%%%%%%%%%%%%%%%%%%%%%%%%%%%%%%%%

\begin{abstract}
We report on the fabrication and electrical characterization of both single layer graphene micron-sized devices and nanoribbons on a hexagonal boron nitride substrate. We show that the micron-sized devices have significantly higher mobility and lower disorder density compared to devices fabricated on silicon dioxide substrate in agreement with previous findings. The transport characteristics of the reactive-ion-etched graphene nanoribbons on hexagonal boron nitride, however, appear to be very similar to those of ribbons on a silicon dioxide substrate. We perform a detailed study in order to highlight both similarities as well as differences. Our findings suggest that the edges have an important influence on transport in reactive ion-etched graphene nanodevices.
\end{abstract}

% 71.15.Mb - condensed matter
% 81.05.ue - materials, carbon based materials, graphene
% 72.80.Vp - electronic transport in graphene

\maketitle
\newpage

Graphene has received tremendous attention since it became easily available for experiments~\cite{Novoselov2004}. Only a few years later, the first graphene-based nanostructures were reported. Among others, small stripes called nanoribbons~\cite{Han2007,Chen2007} and quantum dots~\cite{Stampfer2008} were investigated. In order to understand the limitations of graphene nanostructures better, a large effort was put into studying graphene nanoribbons~\cite{Droescher2011,Oostinga2010,Bai2010,Han2007,Liu2009,Stampfer2009,Todd2009,Gallagher2010,Molitor2010,Han2010,Molitor2009,Terres2011,Chen2007,Lin2008,Hettmansperger2012,Hwang2012} which are the simplest nanostructures and also the building blocks of more complex systems. It was found that charge carriers get localized inside these structures uncontrollably due to either bulk or edge disorder (or a combination of both) which limits reproducibility as well as control of graphene nanodevices~\cite{Todd2009,Stampfer2009,Molitor2009,Liu2009,Han2010,Droescher2011}. 

The electronic transport properties of micron-sized graphene devices were improved by exchanging the traditional silicon dioxide substrate with a more suitable material. The substrate could either be totally removed~\cite{Bolotin2008} (''suspended``) or replaced by hexagonal boron nitride~\cite{Dean2010}. It was shown that suspended graphene in combination with current annealing can lead to high-quality nanostructures~\cite{Tombros2011}. Nanostructures patterned on a substrate are however more amenable to the realization of complex device geometries.

In this paper we investigate if the electrical transport properties change for reactive-ion-etched graphene nanodevices by fabricating them on hexagonal boron nitride rather than on a silicon dioxide substrate. We report on the fabrication and electrical measurements of both micron-sized stripes and nanoribbons built of single layer graphene on a hexagonal boron nitride substrate. The experiment is designed such that a direct comparison with Ref.~\cite{Droescher2011} is possible where similar ribbons fabricated with the same process were investigated on a silicon dioxide substrate. We show that our micron-sized devices exhibit superior transport properties to those on silicon dioxide in agreement with literature~\cite{Dean2010}. At the same time the electrical properties of the nanoribbons on boron nitride are found to be comparable to devices on silicon dioxide. We conclude that the rough edges~\cite{Bischoff2011} in our devices have an important influence on electronic transport in our graphene nanostructures and that simply exchanging the substrate yields no clear improvement.

The investigated devices were fabricated by micromechanical cleaving~\cite{Novoselov2004} of commercial hexagonal boron nitride (hBN; ''PolarTherm Boron Nitride Powder Grade PT110'') and deposition onto Si-SiO$_\mathrm{2}$ chips (285~nm oxide). The hBN flake chosen for the device presented in this work is 13~nm thick as determined by scanning force microscopy. The flake was annealed at 350$^\circ$~C for 3~hours~\cite{Annealing}. A mechanically exfoliated single layer graphene flake was then transferred onto the hBN with a process similar to the one developed by  Ref.~\onlinecite{Dean2010}. Next the resist was removed~\cite{Resist} and the chip was annealed at 330$^\circ$~C for 7~hours~\cite{Annealing}. Contacts were written by electron beam lithography (EBL), developed, metal was evaporated (0.4~nm of chromium followed by 50~nm of gold) and lifted off~\cite{Resist}. We observed that in order to obtain mechanically and thermally stable contacts with low resistance (well below 1~k$\Omega$), the choice of process parameters is crucial: We obtain reproducible results when using short molecule PMMA 50K as a lower resist layer for EBL (and an upper layer of PMMA 950K) together with a sub-nanometer thin sticking-layer of chromium. An additional EBL step was performed and the micron-sized devices were etched out of the flake by reactive ion etching (RIE; mostly argon, few percent of oxygen). Subsequently the resist was removed~\cite{Resist}, the chip was annealed at 320$^\circ$~C for 20~hours~\cite{Annealing} and the micron-sized devices were measured (red curves in Fig.~1b,c, ''before 2nd etching``). After measuring, an additional EBL and RIE step were performed in order to etch the nanoribbons, the resist was removed~\cite{Resist} and the chip annealed at 320$^\circ$~C for 9~hours~\cite{Annealing}. After the second etching step, the device consisted of 3 bulk graphene devices (left as a control group -- see Fig.~1) and 7 nanoribbons (see Figs.~2,3). Representative scanning force images are shown in Fig.~1a and Figs.~2a,b respectively.

In the following, we focus on the electrical measurements of our devices. All measurements presented in this paper are recorded at a temperature of 4.2 K unless otherwise stated and in two-terminal geometry. In all the measurements it was taken care that leakage currents from the back-gate are negligible. Figures~1b,c show the resistance (conductance) versus back-gate voltage of one of the micron-sized devices before (solid) and after (dashed) the second etching step (control group -- unchanged geometry)~\cite{comment2}. A small degradation of the  charge carrier mobility from about 55'000~$\mathrm{cm}^2\mathrm{V}^{-1}\mathrm{s}^{-1}$ to 45'000~$\mathrm{cm}^2\mathrm{V}^{-1}\mathrm{s}^{-1}$ (both lower bounds) at a carrier density of 10$^{10}$~cm$^{-2}$ is found after the second etching step for the control group. The disorder density~\cite{Du2008} is in both cases clearly below 10$^{10}$~cm$^{-2}$ and the charge neutrality point is around 2~V in back-gate voltage indicating small overall doping levels. Figure~1d shows the conductance as a function of back-gate at different magnetic fields. Quantum Hall plateaus are visible down to B~$\approx$~1~T and also the $\nu$~=~1 broken symmetry state is seen. The overswings of the conductance at the edge of the plateaus are due to the device geometry~\cite{Abanin2008}. The charge carrier density extracted from the quantum Hall measurement is in good agreement with values predicted from a parallel plate capacitor model. All three micron-sized devices show similar transport properties that are superior compared to devices on silicon dioxide (compare also Ref.~\onlinecite{Dean2010}) but do not yet reach the quality of the best devices reported in literature (compare for example Ref.~\cite{Dean2011}).
 
% note: all data for Stripe 10-12 
\begin{figure}[tbp]
	\centering\includegraphics[width=0.5\columnwidth]{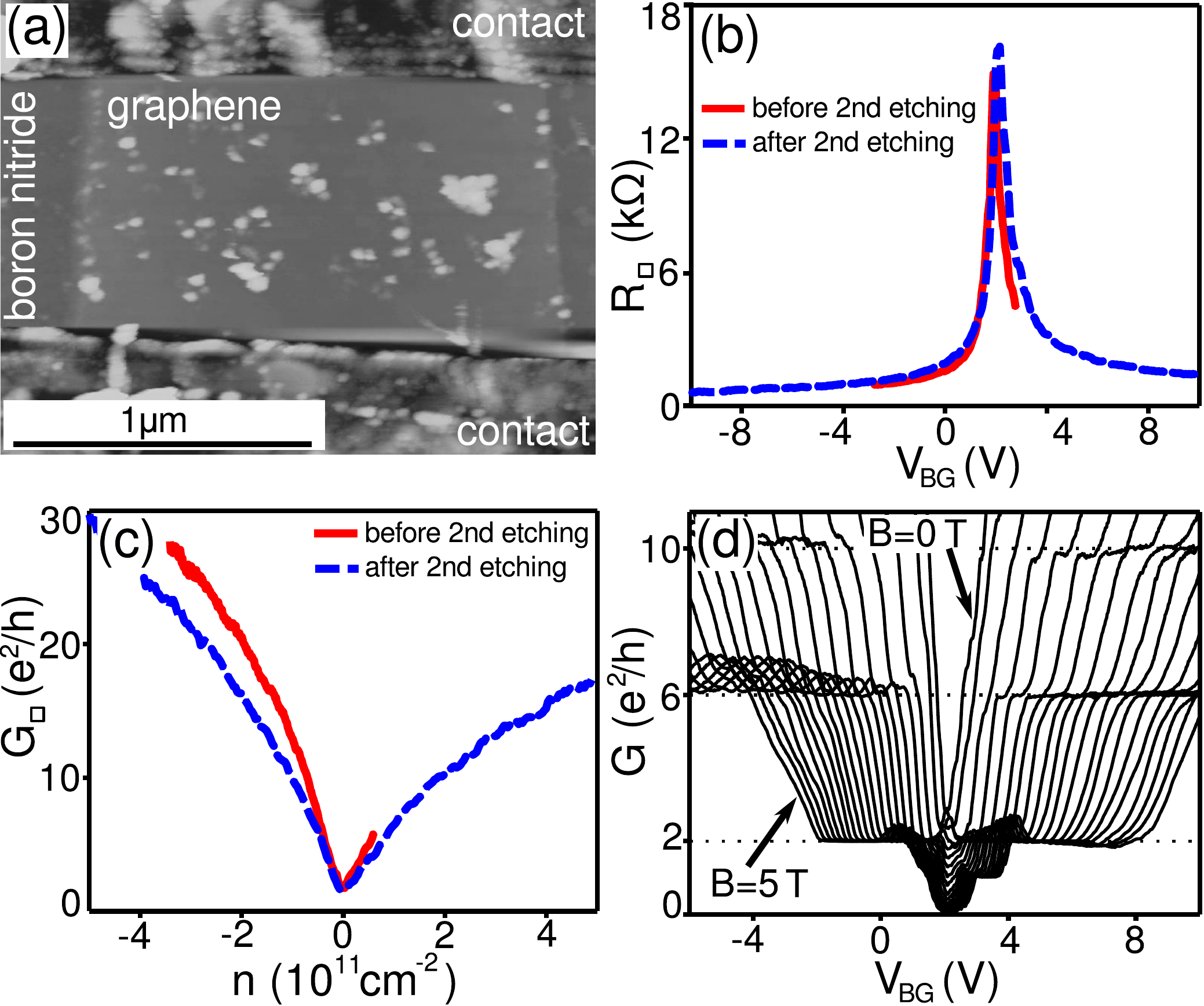}
	\caption{(color online) (a) Scanning force microscopy image (topography) of one of the three micron-sized single layer graphene devices. The graphene is locally flat but some residues from fabrication are  visible (white blobs -- most probably PMMA). (b) Sheet resistance ($R_{\square} = \left[(U/I-R_{\mathrm contact})\times W/L\right]$) of the device in [a] measured as a function of back-gate voltage (contact resistances subtracted). The red solid curve was obtained in a first cooldown where the sample consisted of only micron-sized devices. The blue dashed curve is from a second cooldown for the same micron-sized stripe (not etched - control group) where most of the other bulk devices were etched into nanoribbons. (c) Same data as in [b] but plotted as sheet conductance: $G_{\square} = 1/R_{\square}$. (d) Two-point conductance as a function of back-gate and perpendicular magnetic field (contact resistance of less than 1k$\Omega$ subtracted).}
\end{figure}

Next we will discuss the electrical measurements of the graphene nanoribbons. Out of the seven fabricated ribbons, three didn't show any visible contaminations on the surface when imaged with a scanning force microscope (similar to Fig.~2a) whereas on four some residues from fabrication were visible (similar to Fig.~2b). Figure~2c shows exemplarily the conductance of three ribbons of similar dimensions as a function of back-gate voltage. In all three curves a region of suppressed conductance is visible. The size of the region of suppressed conductance as well as the details of the traces vary strongly for the different ribbons despite their similar lengths and widths. We didn't find any systematic difference between ribbons with residues on the surface (e.g. \#1, \#7 ) and supposedly clean ones (e.g. \#5). In Fig.~2d the dependence of the differential conductance as a function of back-gate voltage and source-drain bias is shown for ribbon \#1: Coulomb diamonds are resolved -- some of them overlapping and some closing at zero bias. These findings are qualitatively similar for all seven investigated ribbons. Such behavior has also been reported for etched graphene nanoribbons on silicon dioxide~\cite{Droescher2011,Oostinga2010,Bai2010,Han2007,Liu2009,Stampfer2009,Todd2009,Gallagher2010,Molitor2010,Han2010,Molitor2009,Terres2011,Lin2008} as well as on suspended devices~\cite{Ki2012}.

\begin{figure}[tbp]
	\centering\includegraphics[width=0.5\columnwidth]{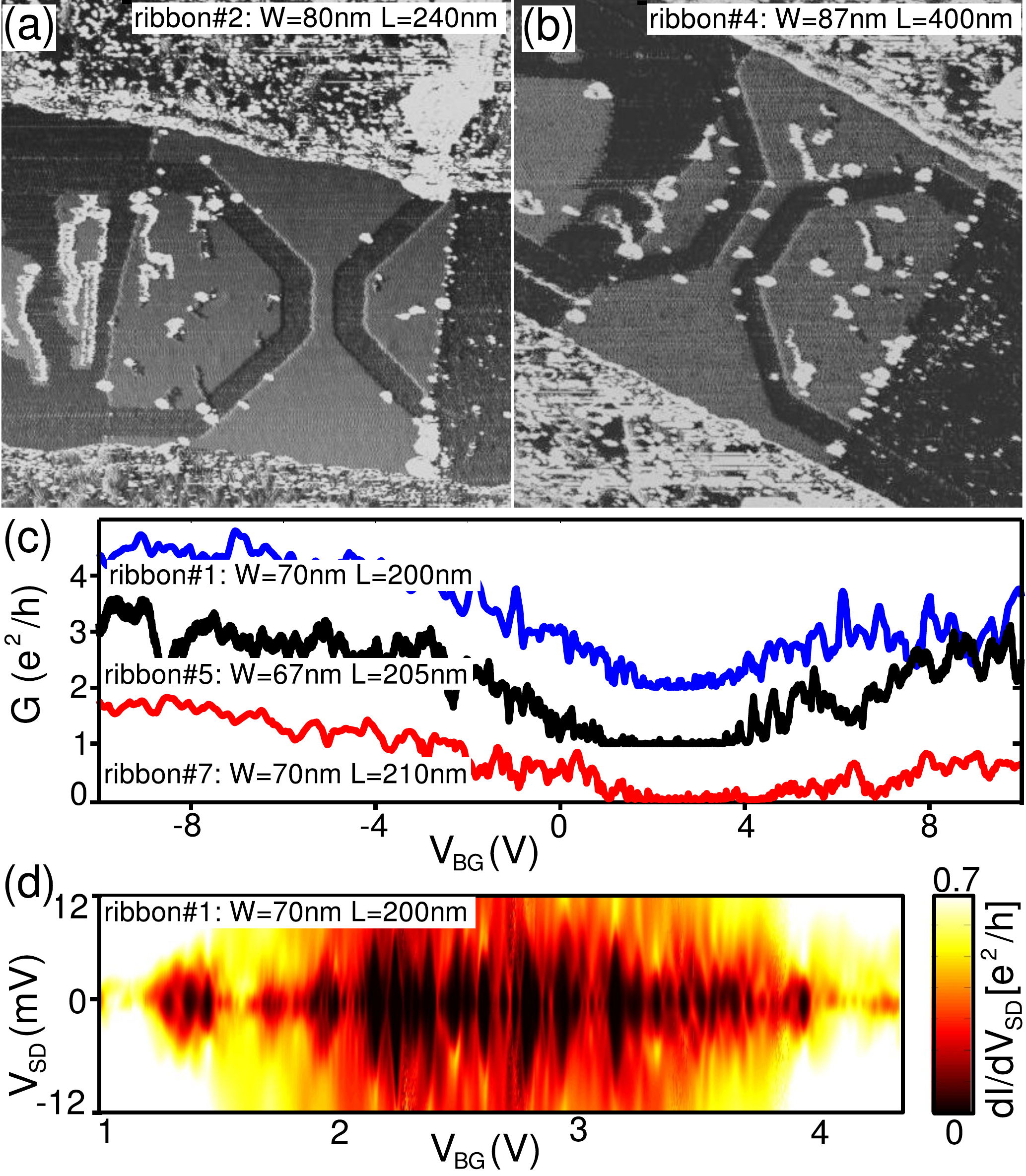}
	\caption{(color online) (a,b) Scanning force microscopy image (phase) of two representative nanoribbons. (c) Conductance of three different ribbons that have comparable dimensions (1~mV bias applied). The curves are vertically offset by e$^2$/h for clarity. (d) Differential conductance as a function of symmetric source-drain bias and back-gate voltage for the blue ribbon trace (ribbon\#1) in [c].}
\end{figure}

A direct comparison of our graphene ribbons on hexagonal boron nitride with ribbons on silicon dioxide is possible with Refs.~\onlinecite{Droescher2011,Terres2011,Molitor2009} due to similar ribbon dimensions and fabrication methods. We find that for all seven ribbons the size of the region of suppressed conductance in back-gate (``back-gate gap'' or ``transport gap``) is similar to the values found in Refs.~\onlinecite{Droescher2011,Molitor2009} within the large experimental variations from ribbon to ribbon. The quite high variation of the transport gap for ribbons of similar dimensions suggests that microscopic details play an important role for this quantity. The region of suppressed conductance in bias (``source-drain gap'') is commonly defined by the extent of the largest Coulomb diamond in source-drain voltage direction~\cite{Droescher2011,Molitor2010,Todd2009,Stampfer2009,Liu2009,Bai2010,Ki2012,Oostinga2010,Terres2011,Molitor2009,Han2010}. For an individual puddle of localized charge, the size of the largest diamond corresponds to the smallest area of localized charge. We find differences in the size of the largest diamond-like feature for ribbons of similar dimensions. The largest diamond usually appears in a region where individual diamonds overlap. We find that in all six ribbons with lengths of around 200~nm and widths ranging from 65~nm to 80~nm, the largest diamond-like feature is larger than 10~meV. These values are comparable or slightly larger than those found in Ref.~\onlinecite{Droescher2011} and significantly larger than those found in Refs.~\onlinecite{Terres2011,Molitor2010,Han2010} for graphene nanoribbons on silicon dioxide (Refs.~\onlinecite{Molitor2010,Han2010} are only partially comparable due to different sizes of the ribbons). The dielectric constant of silicon dioxide and hexagonal boron nitride that enter the charging energy are similar.

Figure~3a shows exemplarily the temperature dependence of the conductance in the region of suppressed conductance for ribbon\#6 (black curves). Figure~3b shows the corresponding finite bias spectroscopy at 4.2 K. Following Refs.~\onlinecite{Han2010,Droescher2011}, the temperature dependence for each point in back-gate is fitted to $$G(T) = G_\mathrm{0}\times \exp\left(\frac{-E_\mathrm{a}}{k_\mathrm{B}\times T}\right)+G_\mathrm{off}~.$$ The values obtained for the activation energy $E_\mathrm{a}$ (blue), prefactor $G_0$ (green) and the ''offset conductance`` $G_\mathrm{off}$ (red) are plotted on top of the data in Figs.~3a,b. The fit is only shown for regions in back-gate where the conductance changes with temperature by at least a factor of three in order to ensure that the fit is meaningful. Like in Ref.~\onlinecite{Droescher2011}, the activation energy fits well to the size of the Coulomb diamonds and $G_\mathrm{off}$ generally overlaps with the lowest temperature curve. In some cases however $G_\mathrm{off}$ has a slightly higher value than the curve at lowest temperature indicating a limitation of the employed model. These findings are also valid for the other investigated ribbons with the tendency that thinner ribbons have lower values of $G_\mathrm{off}$ and higher values of $E_\mathrm{a}$. The temperature dependence of the conductance for our ribbons on hexagonal boron nitride agrees well with ribbons on silicon dioxide\cite{Droescher2011}.

In Fig.~3c the same measurement as in Fig.~3a is shown but at an earlier stage during the same cooldown. The curves are reproducible over short time scales of hours, but the detailed shapes, amplitudes and positions of conductance resonances change on the scale of days. This behavior has been characteristic for all investigated ribbons. Changes can be accelerated by sweeping the back gate over large ranges on the scale of several volts. Also for different cooldowns, the qualitative picture is the same while details vary significantly (ribbons were not exposed to air between cooldowns).

\begin{figure}[tbp]
	\centering\includegraphics[width=0.5\columnwidth]{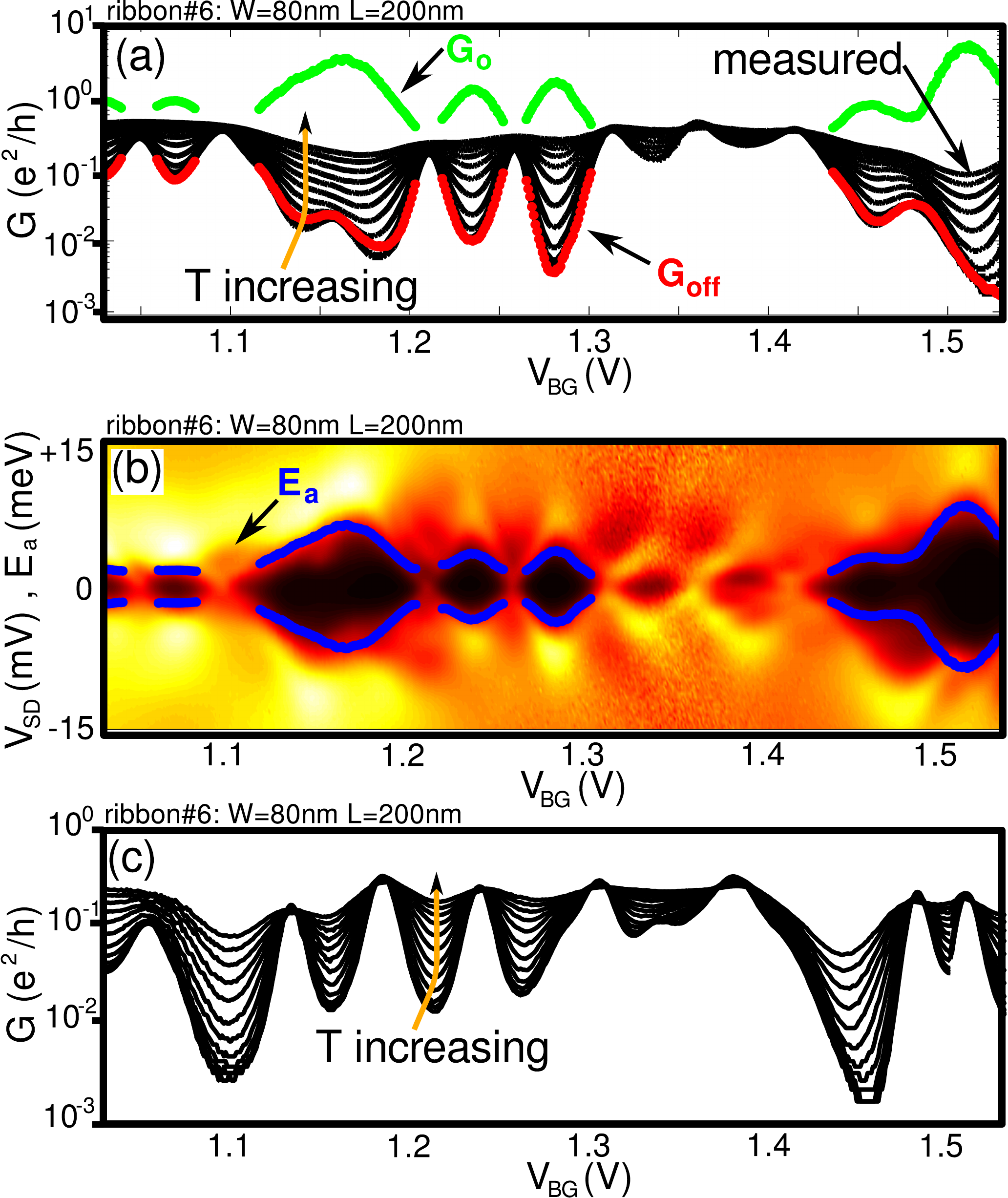}
	\caption{(color online) (a) Temperature dependent measurements of ribbon\#6 in the region of suppressed conductance. The lowest black curve was recorded at a temperature of 4.2~K. The uppermost black curve corresponds to 25~K. The curves were recorded with zero DC bias and 0.35~mV AC modulation to prevent heating. (b) Differential conductance for the same region in back-gate. (c) Same measurement as in [a] but 4~days earlier.}
\end{figure}

In one of the seven ribbons (ribbon\#2) we observe a fully quantized quantum Hall plateau at $\nu=-2$ and a not fully quantized plateau at $\nu=2$ for $B$-field values above 3 T (similar to  Ref.~\onlinecite{Ki2012}). This allows us to determine the capacitance per area of the ribbon which is found to be 21.6~$\mathrm{nF}\mathrm{cm}^{-2}$ or about 1.9 times larger than for the bulk. This enhanced value compared to the plate capacitor model is due to fringe fields that start to play an important role when the thickness of the dielectric is larger than the width of the structure~\cite{Lin2008}. The measured capacitance corresponds to about 26 additional charge carriers that get loaded onto the area of the ribbon for each volt in back-gate. This is in agreement with the roughly 20 Coulomb diamonds that are observed in experiment for a change of one volt in back-gate. Another quantity that can be investigated in more detail as the capacitance of the ribbon relative to the back gate is known, is the charging energy of individual Coulomb diamonds. A direct conversion ($E_\mathrm{C}=e^2/C$) using the plate capacitor model corrected by a factor of 1.9 results in localization areas several times bigger than the ribbon area. This indicates that the capacitance of a puddle of localized charges is enhanced by capacitive coupling to neighboring puddles and leads (see also Ref.~\cite{SusannePHD}).

An additional control experiment was performed after finishing with the measurements descibed above: The devices were covered with PMMA which was subsequently removed again partially with acetone~\cite{CommentControl}. As expected, the PMMA residues degrade the transport properties of the bulk devices (lower mobility of roughly 10'000~$\mathrm{cm}^2\mathrm{V}^{-1}\mathrm{s}^{-1}$, higher disorder density). Compatible with the findings before we however can observe neither a qualitative nor a quantitative change of the transport properties of the nanoribbons.

In conclusion we have shown a clear improvement of both the mobility as well as the disorder density for micron-sized graphene devices on hBN compared to the devices on silicon dioxide in agreement with previous studies~\cite{Dean2010}. Surprisingly no major differences were observed for our reactive ion-etched graphene nanoribbons on hBN compared to their counterparts on silicon dioxide. In both cases, electrical transport is characterized by the random formation of localized charge ''puddles`` and activated transport through them. From these findings we conclude that the edges -- which are expected to be similar for reactive ion-etched ribbons on silicon dioxide as well as on hexagonal boron nitride -- play an important role for charge transport in graphene nanostructures. It is also worth mentioning that the reactive ion etching damages the top-layers of the boron nitride substrate which may additionally influence transport. In order to change the edges of graphene nanostructures in the future, they could for example be passivated~\cite{Kato2011} or controlled on an atomical scale~\cite{Cai2010}. Alternative ways to confine electrons in graphene by locally opening a bandgap include among others functionalizing the graphene~\cite{Jung2008,Robinson2010,Elias2009} or local electrostatic double-gating of bilayer graphene~\cite{Allen2012,Goossens2012}.

Acknowledgements -- We thank J. D. Sanchez-Yamagishi, C. R. Dean, A. F. Young, R. Gorbachev, P. Simonet and A. Varlet for helpful discussions. We thank Momentive Performance Materials for providing the boron nitride. We further thank C. Barengo, I. Altdorfer and P. Studerus for technical support. Financial support by the National Center of Competence in Research on “Quantum Science and Technology“ (NCCR QSIT) is gratefully acknowledged.

%%%%%%%%%%%%%%%%%%%%%%%%%%%%%%%%%%%%%%%%%%%%%%%
%%% bibliography %%%%%%%%%%%%%%%%%%%%%%%%%%%%%%
%%%%%%%%%%%%%%%%%%%%%%%%%%%%%%%%%%%%%%%%%%%%%%%


\begin{thebibliography}{99}
\bibitem{Annealing}
The annealing was performed in a quartz-tube furnace at a constant temperature and constant gas flow of around 2 l/min Ar and 0.2 l/min H$_2$ at ambient pressure.
\bibitem{Resist}
Resist was removed by immersing the chip in 50$^\circ$~C acetone for longer than 30 min and removing the acetone with isopropyl-alcohol.
\bibitem{comment2}
Due to leakage only a small range could be measured in the first cooldown (red solid curves).
\end{thebibliography}
\end{document}